\documentclass[runningheads]{llncs}
\usepackage{graphicx}
\usepackage{subfigure}
\usepackage{multirow}
\usepackage{enumitem}
\usepackage{comment}
\usepackage{caption}
\usepackage{amsmath}
\usepackage{url}

\begin{document}
\title{On the Coherence of Fake News Articles}
%
%
\author{Iknoor Singh\inst{1} 
\and
Deepak P\inst{2} 
\and
Anoop K\inst{3} 
}
%
%
\institute{University of Sheffield, UK \email{iknoor.ai@gmail.com} \and
Queen's University Belfast, UK \email{deepaksp@acm.org} \and
University of Calicut, India \email{anoopk\textunderscore dcs@uoc.ac.in}
}
\maketitle              
\begin{abstract}
The generation and spread of fake news within new and online media sources is emerging as a phenomenon of high societal significance. Combating them using data-driven analytics has been attracting much recent scholarly interest. In this computational social science study, we analyze the textual coherence of fake news articles vis-a-vis legitimate ones. We develop three computational formulations of textual coherence drawing upon the state-of-the-art methods in natural language processing and data science. Two real-world datasets from widely different domains which have fake/legitimate article labellings are then analyzed with respect to textual coherence. We observe apparent differences in textual coherence across fake and legitimate news articles, with fake news articles consistently scoring lower on coherence as compared to legitimate news ones. While the relative coherence shortfall of fake news articles as compared to legitimate ones form the main observation from our study, we analyze several aspects of the differences and outline potential avenues of further inquiry. 

\end{abstract}
\section{Introduction}
\label{sec:intro}

The spread of {\it fake news} is increasingly being recognized as a global issue of enormous significance. The phenomenon of fake news, or disinformation disguised as news, started gaining rampant attention around the 2016 US presidential elections~\cite{allcott2017social}. While politics remains the domain which attracts most scholarly interest in studying the influence of fake news~\cite{davies2016age}, the impact of alternative facts on economic~\cite{hopkin2018post} and healthcare~\cite{speed2017rise} sectors are increasingly getting recognized. Of late, the news ecosystem has evolved from a small set of regulated and trusted sources to numerous online news sources and social media. Such new media sources come with limited liability for disinformation, and thus are easy vehicles for fake news. Data science methods for fake news detection within social media such as Twitter has largely focused on leveraging the social network and temporal propagation information such as response and retweet traces and their temporal build-up; the usage of the core content information within the tweet has been shallow. In fact, some recent techniques (e.g.,~\cite{wu2018tracing}) achieve state-of-the-art performance without using any content features whatsoever. The task landscape, however, changes significantly when one moves from the realm of tweets to online news sources (such as those in Wikipedia's list\footnote{https://en.wikipedia.org/wiki/List\_of\_fake\_news\_websites}); the latter form a large fraction of fake news that are debunked within popular debunking sites such as Snopes\footnote{http://www.snopes.com}. Fake news within online news sources (e.g.,~\cite{indonesian}) are characterized by scanty network and propagation information; this is so since they are typically posted as textual articles within websites (as against originating from social media accounts), and their propagation happens on external social media websites through link sharing. Further, it is often necessary to bank on exploiting text information in order to develop fake news detection methods for scenarios such as those of a narrow scope (e.g., a local council election, or regional soccer game) even for Twitter, since the narrow scope would yield sparse and unreliable propagation information. Accordingly, there has been some recent interest in characterizing fake news in terms of various aspects of textual content. Previous work along this direction has considered satirical cues~\cite{rubin2016fake}, expression of stance~\cite{chopra2017towards}, rhetorical structures~\cite{rubin2015towards} and topical novelty~\cite{vosoughi2018spread}. 

In this computational social science study, we evaluate the textual coherence of fake news articles vis-a-vis legitimate ones. While our definitions of textual coherence will follow in a later section, it is quite intimately related to the notion of {\it cohesion} and {\it coherence} in language studies~\cite{morris1991lexical}. We choose to use the term {\it coherence} due to being more familiar to the computing community. Cohesion has been considered as an important feature for assessing the structure of text~\cite{morris1991lexical} and has been argued to play a role in writing quality~\cite{alotaibi2015role,mcculley1985writing,crossley2011text}. 

\section{Related Work}

With our study being on coherence on fake news as assessed using their textual content, we provide some background into two streams of literature. First, we briefly summarize some recent literature around fake news with particular emphasis to studies directed at their lexical properties and those directed at non-lexical integrity and coherence. Second, we outline some natural language processing (NLP) techniques that we will use as technical building blocks to assess textual coherence.

\subsubsection*{Characteristics of Fake News:}


In the recent years, there have been abundant explorations into understanding the phenomenon of fake news. We now briefly review some selected work in the area. In probably the first in-depth effort at characterizing text content,~\cite{rubin2015towards} make use of rhetorical structure theory to understand differences in distribution of rhetorical relations across fake and legitimate news. They identify that disjunction and restatement appear much more commonly within legitimate stories. Along similar lines,~\cite{potthast2017stylometric} perform a stylometric analysis inquiring into whether writing style can help distinguish hyperpartisan news from others, and fake news from legitimate ones (two separate tasks). While they identify significant differences in style between hyperpartisan and non-hyperpartisan articles, they observe that such differences are much more subdued across fake and legitimate news and are less likely to be useful for the latter task. In an extensive longitudinal study~\cite{vosoughi2018spread} to understand the propagation of fake and legitimate news, the authors leverage topic models~\cite{steyvers2007probabilistic} to quantify the divergence in character between them. They find that human users have an inherent higher propensity to spread fake news and attribute that propensity to users perceiving fake news as more novel in comparison to what they have seen in the near past, novelty assessed using topic models. Conroy et al.~\cite{conroy2015automatic} summarize the research into identification of {\it deception cues} (or tell-tale characteristics of fake news) into two streams; viz., linguistic and network. While they do not make concrete observations on the relative effectiveness of the two categories of cues, they illustrate the utility of external knowledge sources such as Wikipedia in order to evaluate veracity of claims. Moving outside the realm of text analysis, there has been work on quantifying the differences between fake and legitimate news datasets in terms of image features~\cite{jin2017novel}. They observe that there are often images along with microblog posts in Twitter, and these may hold cues as to the veracity of the tweet. With most tweets containing only one image each, they analyze the statistical properties of the dataset of images collected from across fake tweets, and compare/contrast them with those from a dataset of images from across legitimate tweets. They observe that the average visual coherence, i.e., the average pairwise similarities between images, is roughly the same across fake and legitimate image datasets; however, the fake image dataset has a larger dispersion of coherence scores around the mean. In devising our lexical coherence scores that we use in our study, we were inspired by the formulation of visual coherence scores as aggregates of pairwise similarities. 

\subsubsection*{NLP Building Blocks for Coherence Assessments:}

In our study, we quantify lexical coherence by building upon advancements in three different directions in natural language processing. We briefly outline them herein. 

\subsubsection{Text Embeddings} 

Leveraging a dataset of text documents to learn distributional/vector representations of words, phrases and sentences has been a recent trend in natural language processing literature, due to pioneering work such as word2vec~\cite{mikolov2013distributed} and GloVe~\cite{pennington2014glove}. Such techniques, called word embedding algorithms, map each word in the document corpus to a vector of pre-specified dimensionality by utilizing the lexical proximity of words within the documents in which they appear. Thus, words that appear close to each other often within documents in the corpus would get mapped to vectors that are proximal to each other. These vectors have been shown to yield themselves to meaningful algebraic manipulation~\cite{mikolov2013linguistic}.
While word embeddings are themselves useful for many practical applications, techniques for deriving a single embedding/vector for larger text fragments such as sentences, paragraphs and whole documents~\cite{le2014distributed} have been devised. That said, sentence embeddings/vectors formed by simply averaging the vectors corresponding to their component words, often dubbed average word2vec (e.g.,~\cite{dilawar2018understanding}), often are found to be competitive with more sophisticated embeddings for text fragments. In our first computational model for lexical coherence assessment, we will use average word2vec vectors to represent sentences. 

\subsubsection{Explicit Semantic Analysis} 

Structured knowledge sources such as Wikipedia encompass a wide variety of high-quality manually curated and continuously updated knowledge. Using them for deriving meaningful representations of text data has been a direction of extensive research. A notable technique~\cite{gabrilovich2007computing} along this direction attempts to represent text as vectors in a high dimensional space formed by Wikipedia concepts. Owing to using Wikipedia concepts explicitly, it is called explicit semantic analysis. In addition to generating vector representations that is then useful for a variety of text analytics tasks, these vectors are intuitively meaningful and easy to interpret due to the dimensions mapping directly to Wikipedia articles. This operates by processing Wikipedia articles in order to derive an inverted index for words (as is common in information retrieval engines~\cite{ounis2005terrier}); these inverted indexes are then used to convert a text into a vector in the space of Wikipedia dimensions. Explicit Semantic Analysis, or ESA as it is often referred to, has been used for a number of different applications including those relating to computing semantic relatedness~\cite{scholl2010extended}. In our second computation approach, we will estimate document lexical coherence using sentence-level ESA vectors. 

\subsubsection{Entity Linkings} 

In the realm of news articles, entities from knowledge bases such as Wikipedia often get directly referenced in the text. For example, the fragments {\it UK} and {\it European Union} in {\it `UK is due to leave the European Customs Union in 2020'} may be seen to be intuitively referencing the respective entities, i.e., {\it United Kingdom} and {\it European Union}, in a knowledge base. Entity Linking~\cite{hoffart2011robust} ({\it aka} named entity disambiguation) methodologies target to identify such references, effectively establishing a method to directly link text to a set of entities in a knowledge base; Wikipedia is the most popular knowledge base used for entity linking (e.g.,~\cite{mihalcea2007wikify}). Wikipedia entities may be thought of as being connected in a graph, edges constructed using hyperlinks in respective Wikipedia articles. Wikipedia graphs have been used for estimating semantic relatedness of text in previous work~\cite{yeh2009wikiwalk}. In short, converting a text document to a set of Wikipedia entities referenced within them (using entity linking methods) provides a semantic grounding for the text within the Wikipedia graph. We use such Wikipedia-based semantic representations in order to devise our third computational technique for measuring lexical coherence. 

\section{Research Objective and Computational Framework}

\subsubsection*{Research Objective:} 

Our core research objective is to study whether fake news articles differ from legitimate news articles along the dimension of {\it  coherence as estimated using their textual content} using computational methods. We use the term coherence to denote the overall consistency of the document in adhering to core focus theme(s) or topic(s). In a way, it may be seen as contrary to the notion of dispersion or scatter. A document that switches themes many times over may be considered as one that lacks coherence. 

\subsection{Computational Framework for Lexical Coherence} 

For the purposes of our study, we need a computational notion of coherence. Inspired by previous work within the realm of images where visual coherence is estimated using aggregate of pairwise similarities~\cite{jin2017novel}, we model the coherence of an article as the average/mean of pairwise similarities between 'elements' of the article. Depending on the computational technique, we model elements as either sentences or Wikipedia entities referenced in the article. For the sentence-level structuring, for a document $D$ comprising $d_n$ sentences $\{ s_1, s_2, \ldots, s_{d_n}\}$, coherence is outlined as:

\begin{equation}
C_{sent}(D) = \mathrm{mean}\bigg\{ \mathrm{sim}(rep(s_i),rep(s_j)) | s_i,s_j \in D, i \neq j \bigg\}
\end{equation}
where $rep(s)$ denotes the representation of the sentence $s$, and $\mathrm{sim}(.,.)$ is a suitable similarity function over the chosen representation. Two of our computational approaches use sentence level coherence assessments; they differ in the kind of representation, and consequently the similarity measure, that they use. The third measure uses entity linking to identify Wikipedia entities that are used in the text document. For a document $D$ comprising references to $d_m$ entities $\{ e_1, e_2, \ldots, e_{d_m} \}$, coherence is computed as:

\begin{equation}
C_{ent}(D) = \mathrm{mean}\bigg\{ \mathrm{sim}(e_i,e_j) | e_i,e_j \in D, i \neq j \bigg\}
\label{eq:cent}
\end{equation}
where $\mathrm{sim}(.,.)$ is a suitable similarity measure between pairs of Wikipedia entities. We consistently use numeric vectors as representations of sentences and entities, and employ cosine similarity\footnote{https://en.wikipedia.org/wiki/Cosine\_similarity}, the popular vector similarity measure, to compute similarities between vectors. 


\subsection{Profiling Fake and Legitimate news using Lexical Coherence} 

Having outlined our framework for computing document-specific lexical coherence, we now describe how we use it in order to understand differences between fake and legitimate news articles. Let $\mathcal{F} = \{ \ldots, F, \ldots \}$ and $\mathcal{L} = \{ \ldots, L, \ldots \}$ be separate datasets of fake news and legitimate news articles respectively. Each document in $\mathcal{F}$ and $\mathcal{L}$ is subjected to the coherence assessment, yielding a single document-specific score (for each computational approach for quantifying lexical coherence). These yield separate sets of lexical coherence values for $\mathcal{F}$ and $\mathcal{L}$:

\begin{equation}
C(\mathcal{F}) = \{ \ldots, C(F), \ldots \} \hspace{0.5in} 
C(\mathcal{L}) = \{ \ldots, C(L), \ldots \}
\end{equation}

where $C(.)$ is either of $C_{sent}$ or $C_{ent}$. Aggregate statistics across the sets $C(\mathcal{F})$ and $C(\mathcal{L})$, such as mean and standard deviation, enable quantifying the relative differences in textual coherence between fake and legitimate news. 

\section{Computational Approaches}

We now outline our three computational approaches to quantify coherence at the document level. The first and second approaches use sentence-level modelling, and leverage word embeddings and explicit semantic analysis respectively. The third uses entity linking to convert each document into a set of entities, followed by computing the coherence at the entity set level. Each coherence quantification method is fully specified with the specification of how the vector representation is derived for the elements (entities or sentences) of the document. 

\subsubsection*{Coherence using Word Embeddings}

This is the first of our sentence-level coherence assessment methods. As outlined in the related work section, word2vec~\cite{mikolov2013distributed} is among the most popular word embedding methods. Word2vec uses a shallow, two-layer neural network that is trained to reconstruct linguistic contexts of words over a corpus of documents. Word2vec takes as its input a large corpus of text and produces a vector space, typically of several hundred dimensions, with each unique word in the corpus being assigned a corresponding vector in the space. 
We use a pre-trained word2vec vector dataset that was trained over a huge corpus\footnote{https://github.com/mmihaltz/word2vec-GoogleNews-vectors} since they are likely to be better representations being learnt over a massive dataset. Each sentence $s_i$ in document $D$ is then represented as the average of the word2vec vectors of the words, denoted $\mathrm{word2vec}(w)$, it contains:

\begin{equation}
rep(s_i) = \mathrm{mean}\{ \mathrm{word2vec}(w) | w \in s_i \}
\end{equation}
This completes the specification of coherence quantification using embeddings. 

\subsubsection*{Coherence using Explicit Semantic Analysis}


Explicit Semantic Analysis (ESA)~\cite{gabrilovich2007computing} forms the basis of our second sentence-level coherence quantification method. ESA starts with a collection of text articles sources from a knowledge base, typically Wikipedia; each article is turned into a {\it bag of words}. Each word may then be thought of as being represented as a vector over the set of Wikipedia articles, each element of the vector directly related to the number of times it appears in the respective article. The ESA representation for each sentence is then simply the average of the ESA vectors of the words that it contains. 

\begin{equation}
rep(s_i) = \mathrm{mean}\{ \mathrm{esa}(w) | w \in s_i \}
\end{equation}
where $\mathrm{esa}(w)$ is the vector representation of the word $w$ under ESA. 

\subsubsection*{Coherence using Entity Linking}


Given a document, entity linking (EL) methods~\cite{gupta2017entity} identify mentions of entities within them. Identifying Wikipedia entities to associate references to, is often complex and depends on the context of the word. EL algorithms use a variety of different heuristics in order to accurately identify the set of entities referenced in the document. 
We would like to now convert each entity thus identified to a vector so that it may be used for document coherence quantification within our framework. Towards this, we use the Wikipedia2vec technique~\cite{yamada2018wikipedia2vec} which is inspired by Word2vec and forms vectors for Wikipedia entities by processing the corpus of Wikipedia articles. Thus, the representation of each entity is simply the Wikipedia2vec vector associated with that entity. That representation then feeds into Eq~\ref{eq:cent} for coherence assessments. 



\section{Experimental Study}


\subsection{Datasets}

In order to ensure the generalizability of the insights from our coherence study, we evaluate the coherence scores over two datasets. The first one, {\it ISOT fake news dataset}, is a publicly available dataset comprising 10k+ articles focused on politics. The second dataset is one sourced by ourselves comprising 1k articles on {\it health and well-being} (HWB) from various online sources. These datasets are very different both in terms of size and the kind of topics they deal with. It may be noted that the nature of our task makes datasets comprising long text articles more suitable. Most fake news datasets involve tweets, and are thus sparse with respect to text data\footnote{Tweets are limited to a maximum of 280 characters.}, making them unsuitable for textual coherence studies such as ours. Thus, we limit our attention to the aforementioned two datasets both of which comprise textual articles. We describe them separately herein. 

\subsubsection{ISOT Fake News Dataset}

The ISOT Fake News dataset\footnote{https://www.uvic.ca/engineering/ece/isot/datasets/index.php}~\cite{ahmed2017detection} is the largest public dataset of textual articles with fake and legitimate labellings that we have come across. The ISOT dataset comprises various categories of articles, of which the {\it politics} category is the only one that appears within both {\it fake} and {\it real/legitimate} labellings. Thus, we use the politics subset from both fake and legitimate categories for our study. The details of the dataset as well as the statistics from the sentence segmentation and entity linking appear in Table~\ref{tab:dataset}. 

\subsubsection{HWB Dataset}

We believe that health and well-being is another domain that is often targeted by fake news sources. As an example, fake news on topics such as vaccinations has raised significant concerns~\cite{iacobucci2019vaccination} in recent times, not to mention the COVID-19 pandemic. Thus, we curated a set of articles with fake/legitimate labellings tailored to the health domain. For the legitimate news articles, we crawled 500 news documents on health and well-being from reputable sources such as CNN, NYTimes, Washington Post and New Indian Express. For fake news, we crawled 500 articles on similar topics from well-reported misinformation websites such as BeforeItsNews, Nephef and MadWorldNews. These were manually verified for category suitability, thus avoiding blind reliance on source level labellings. This dataset, which we will refer to as HWB, short for {\it health and well-being}, will be made available at \url{https://dcs.uoc.ac.in/cida/resources/hwb.html}. HWB dataset statistics also appear in Table~\ref{tab:dataset}. 

\subsubsection{On the Article Size Disparity} 

The disparity in article sizes between fake and legitimate news articles is important to reflect upon, in the context of our comparative study. In particular, it is important to note how coherence assessments may be influenced by the number of sentences and entity references within each article. 
It may be intuitively expected that coherence quantification would yield a lower value for longer documents than for shorter ones, given that all pairs of sentences/entities are used in the comparison. 
Our study stems from the hypothesis that fake articles may be {\it less} coherent; the article length disparity in the dataset suggests that the null hypothesis assumption (that coherence is similar across fake and legitimate news) being true would yield {\it higher} coherence score for the fake news documents (being shorter). {\it In a way, it may be observed that any empirical evidence illustrating {\it lower} coherence scores for fake articles, as we observe in our results, could be held to infer a stronger departure from the null hypothesis than in a dataset where fake and legitimate articles were of similar sizes.} Thus, the article length distribution trends only deepen the significance of our results that points to lower coherence among fake articles. 

\begin{table}[t]
\centering
\caption{Dataset Statistics (SD = Standard Deviation)}
\label{tab:dataset}
\begin{tabular}{|c|l|c|c|c|}
\hline
Dataset &
  \multicolumn{1}{c|}{Category} &
  \#Articles &
  \begin{tabular}[c]{@{}c@{}}\#Sentences Per Article \\ Mean (SD)\end{tabular} &
  \begin{tabular}[c]{@{}c@{}}\#Entities Per Article \\ Mean (SD)\end{tabular} \\ \hline
\multirow{2}{*}{ISOT} & Fake       & 5816  & 11.56 (10.54) & 38.99 (30.17) \\
                      & Legitimate & 11175 & 14.65 (10.71) & 46.48 (27.31) \\ \hline
\multirow{2}{*}{HWB}  & Fake       & 500   & 24.62 (18.04) & 49.56 (28.67) \\
                      & Legitimate & 500   & 28.24 (19.37) & 62.91 (37.07) \\ \hline
\end{tabular}
\vspace{-0.3in}
\end{table}

\subsection{Experimental Setup}

We now describe some details of the experimental setup we employed. The code was written in Python. NLTK\footnote{https://www.nltk.org/}, a popular natural language toolkit, was used for sentence splitting and further processing. The word embedding coherence assesments were performed using Google's pre-trained word2vec corpus\footnote{https://code.google.com/archive/p/word2vec/}, which was trained over news articles. Explicit Semantic Analysis (ESA) was run using the publicly available EasyESA\footnote{https://github.com/dscarvalho/easyesa} implementation. For the entity linking method, the named entities were identified using the NLTK toolkit, and their vectors were looked up on the Wikipedia2vec\footnote{https://github.com/wikipedia2vec/wikipedia2vec} pre-trained model\footnote{https://wikipedia2vec.github.io/wikipedia2vec/pretrained/}. 

\subsection{Analysis of Text Coherence}

We first analyze the mean and standard deviation of coherence scores of fake and legitimate news articles as assessed using each of our three methods. Higher coherence scores indicate higher textual coherence. Table~\ref{tab:summary} summarizes the results. {\it In each of the three coherence assessments across two datasets, thus six combinations overall, the fake news articles were found to be less coherent than the legitimate news ones on the average.} The difference was found to be statistically significant with $p<0.05$ under the two-tailed t-test\footnote{https://www.itl.nist.gov/div898/handbook/eda/section3/eda353.htm} in five of six combinations, recording very low p-values (i.e., strong difference) in many cases. 


\begin{table}[t]
\centering
\caption{Coherence Results Summary (SD = Standard Deviation). Statistically signigicant results with $p<0.05$ (two-tailed t-test) in bold. XE-Y is a commonly used mathematical abbreviation to stand for $X\times 10^{-Y}$}
\label{tab:summary}
\resizebox{\textwidth}{!}{%
\begin{tabular}{|c|l|c|c|c|}
\hline
Dataset &
  \multicolumn{1}{c|}{Category} &
  \begin{tabular}[c]{@{}c@{}}Word Embedding \\ Coherence Mean (SD)\end{tabular} &
  \begin{tabular}[c]{@{}c@{}}ESA Coherence\\ Mean (SD)\end{tabular} &
  \begin{tabular}[c]{@{}c@{}}Entity Linking \\ Coherence Mean (SD)\end{tabular} \\ \hline
\multirow{4}{*}{ISOT} & Fake             & 0.546518 (0.0070) & 0.999218 (4.90E-05) & 0.277454 (0.0010)  \\
                      & Legitimate       & 0.567870 (0.0055) & 0.999474 (2.45E-05) & 0.286689 (0.0003)  \\
                      & Difference in \% & 3.91\%            & 0.03\%              & 3.33\%             \\
                      & p-value          & \textbf{6.29E-60} & \textbf{0.013341}   & \textbf{1.90E-100} \\ \hline
\multirow{4}{*}{HWB}  & Fake             & 0.468907 (0.0055) & 0.995245 (0.0040)   & 0.307874 (0.0006)  \\
                      & Legitimate       & 0.506322 (0.0059) & 0.997276 (0.0020)   & 0.318574 (0.0008)  \\
                      & Difference in \% & 7.98\%            & 0.20\%              & 3.48\%             \\
                      & p-value          & \textbf{1.46E-14} & 0.557432            & \textbf{6.27E-10}  \\ \hline
\end{tabular}%
}\end{table}

\subsubsection{Trends across methods}

The largest difference in means are observed for the word embedding coherence scores, with the legitimate news articles being around 4\% and 8\% more coherent than fake news articles in the ISOT and HWB datasets respectively. The lowest p-values are observed for the entity linking method, where the legitimate articles are 3+\% more coherent than fake news articles across datasets for the ISOT dataset; the p-value being in the region of 1E-100 indicates the presence of consistent difference in coherence across fake and legitimate news articles. On the other hand, the coherence scores for ESA vary only slightly in magnitude across fake and legitimate news articles. This is likely because ESA is primarily intended to separate articles from different domains; articles within the same domain thus often get judged to be very close to each other, as we will see in a more detailed analysis in the later section. The smaller differences across fake and legitimate news articles are still statistically significant for the ISOT dataset (p-value $<0.05$), whereas they are not so in the case of the much smaller HWB dataset. It may be appreciated that statistical significance assesments depend on degrees of freedom, roughly interpreted as the number of independent samples; this makes it harder to approach statistical significance in small datasets. Overall, these results also indicate that the word embedding perspective is best suited, among the three methods, to discern textual coherence differences between legitimate and fake news.

\subsubsection{Coherence Spread}

The spread of the coherence across articles was seen to be largely broader for fake news articles, as compared to legitimate ones. This is evident from the higher standard deviations exhibited by the coherence scores in the majority of cases. From observations over the dataset, we find that the coherence scores for legitimate articles generally form a unimodal distribution, whereas the distribution of coherence scores across fake news articles show some minor deviations from unimodality. In particular, we find a small number of scores clustered in the low range of the spectrum, and another much larger set of scores forming a unimodal distribution centered at a score lesser than that the respective centre for the legitimate news articles. This difference in character, of which more details follow in the next section, reflects in the higher standard deviation in coherence scores for the fake news documents. 

\begin{figure}[!htb]
    \centering
        \includegraphics[width=0.5\linewidth]{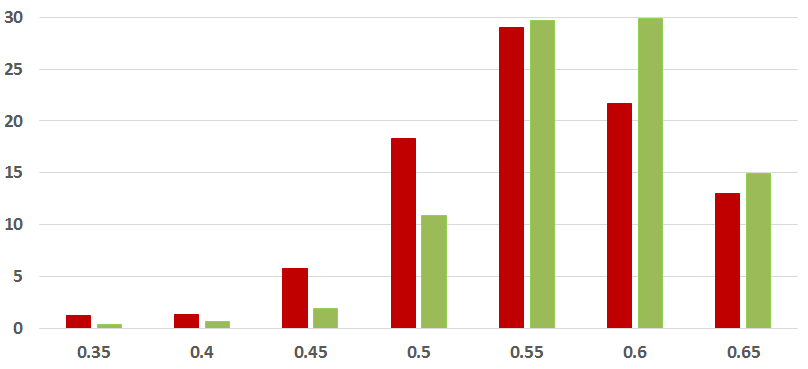}
        \caption{{\bf ISOT Word Embedding Coherence Histogram} (\%articles vs. coherence scores): Legitimate  and Fake articles' scores in green and red respectively.}
        \label{fig:wordembhistisot}
        \vspace{-0.3in}
\end{figure}

\begin{figure}[!htb]
    \centering
        \includegraphics[width=0.5\linewidth]{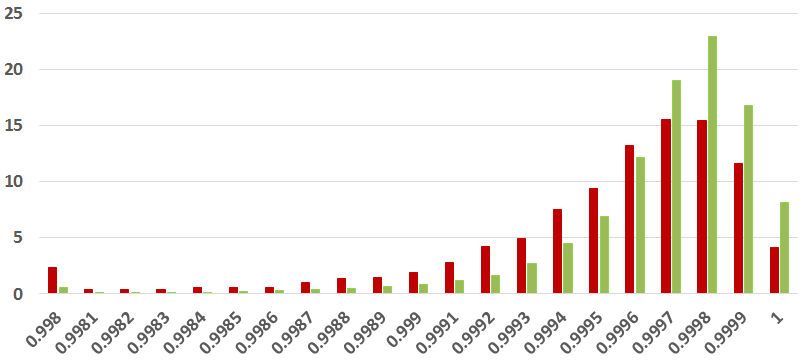}
        \caption{{\bf ISOT ESA Coherence Histogram} (\%articles vs. coherence scores): Legitimate  and Fake articles' scores in green and red respectively.}
        \label{fig:esahistisot}
\end{figure}

\begin{figure}[!htb]
    \centering
        \includegraphics[width=0.5\linewidth]{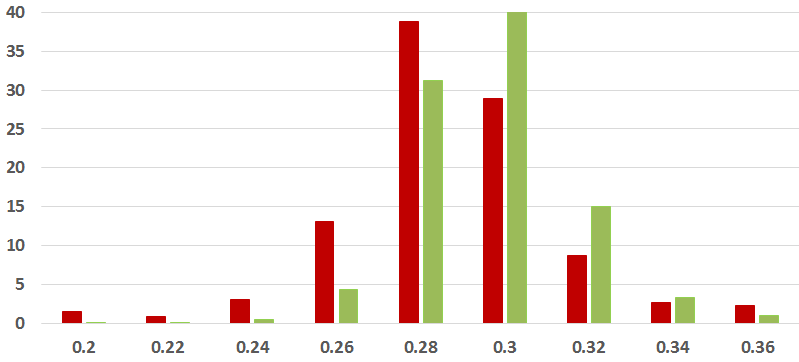}
        \caption{{\bf ISOT Entity Linking Coherence Histogram} (\%articles vs. coherence scores): Legitimate  and Fake articles' scores in green and red respectively.}
        \label{fig:entityhistisot}
        \vspace{-0.2in}
\end{figure}

\subsubsection{Coherence Score Histograms}

\begin{figure}[!htb]
    \centering
        \includegraphics[width=0.5\linewidth]{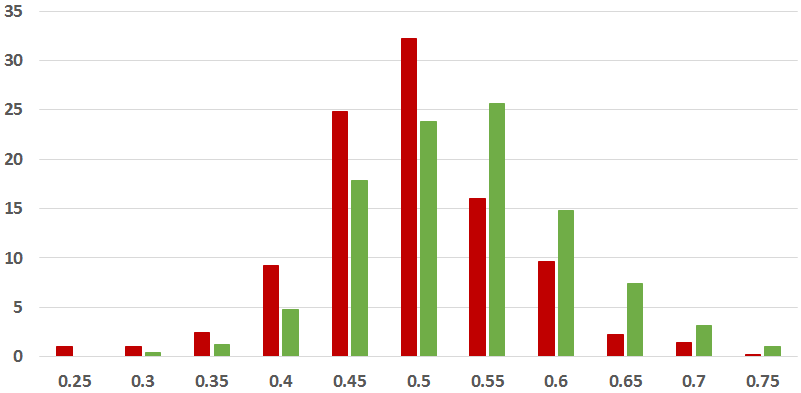}
        \caption{{\bf HWB Word Embedding Coherence Histogram} (\%articles vs. coherence scores): Legitimate  and Fake articles' scores in green and red respectively.}
        \label{fig:wordembhisthwb}
        \vspace{-0.3in}
\end{figure}

\begin{figure}[!htb]
    \centering
        \includegraphics[width=0.5\linewidth]{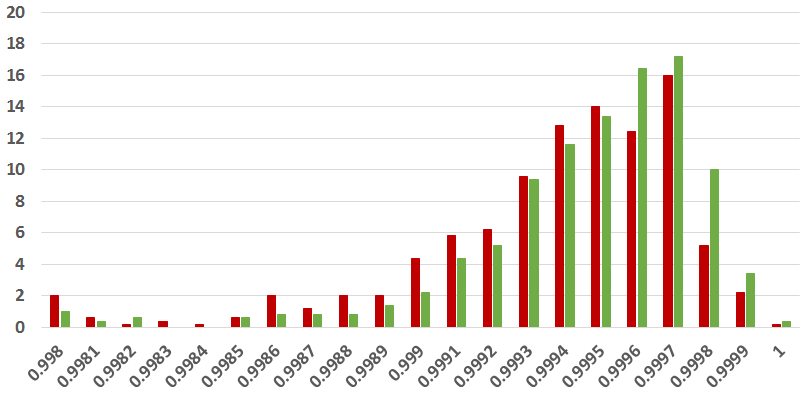}
        \caption{{\bf HWB ESA Coherence Histogram} (\%articles vs. coherence scores): Legitimate  and Fake articles' scores in green and red respectively.}
        \label{fig:esahisthwb}
        \vspace{-0.3in}
\end{figure}

\begin{figure}[!htb]
    \centering
        \includegraphics[width=0.5\linewidth]{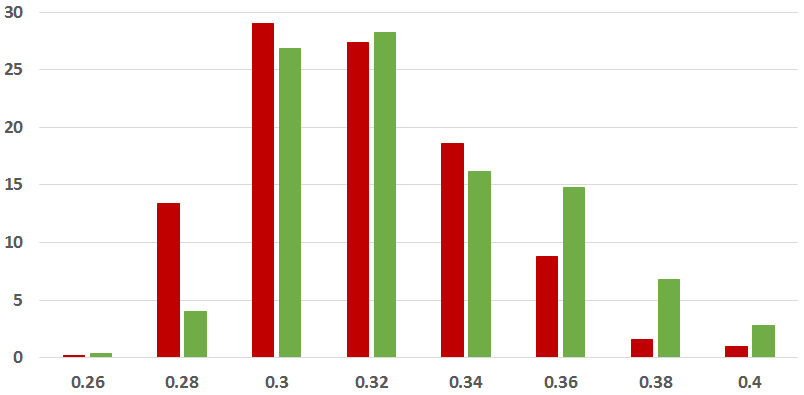}
        \caption{{\bf HWB Entity Linking Coherence Histogram} (\%articles vs. coherence scores): Legitimate  and Fake articles' scores in green and red respectively.}
        \label{fig:entityhisthwb}
        \vspace{-0.3in}
\end{figure}

We visualize the nature of the coherence score distributions for the various methods further by way of histogram plots in Figures~\ref{fig:wordembhistisot},~\ref{fig:esahistisot} and~\ref{fig:entityhistisot} for the ISOT dataset. The corresponding histograms for the HWB dataset appear in Figures~\ref{fig:wordembhisthwb},~\ref{fig:esahisthwb} and~\ref{fig:entityhisthwb}. We have set the histogram buckets in a way to amplify the region where most documents fall, density of scores being around different ranges for different methods. This happens to be around 0.5 for word embedding scores, 0.999 for ESA scores, and around 0.3 for entity linking scores. With the number of articles differing across the fake and legitimate subsets, we have set the Y-axis to indicate the percentage of articles in each of the ranges (as opposed to raw frequency counts), to aid meaningful visual comparison. All documents that fall outside the range in the histogram are incorporated into the leftmost or rightmost pillar in the histogram as appropriate. 

The most striking high-level trend, across the six histograms is as follows. When one follows the histogram pillars from left to right, the red pillar (corresponding to fake news articles) is consistently taller than the green pillar (corresponding to legitimate news articles), until a point beyond which the trend reverses; from that point onwards the green pillar is consistently taller than the red pillar. Thus, the lower coherence scores have a larger fraction of fake articles than legitimate ones and vice versa. For example, this point of reversal is at 0.55 for Figure~\ref{fig:wordembhistisot}, 0.9997 for Figure~\ref{fig:entityhistisot} and 0.9996 for Figure~\ref{fig:esahisthwb}. There are only two histogram points that are not very aligned with this trend, both for the entity linking method, which are 0.36 in Figure~\ref{fig:entityhistisot} and 0.34 in Figure~\ref{fig:entityhisthwb}; even in those cases, the high-level trend is still consistent with our analysis. 

The second observation, albeit unsurprising, is the largely single-peak (i.e., unimodal) distribution of the coherence score in each of the six charts for both fake and legitimate news articles; this also vindicates our choice of the statistical significance test statistic in the previous section, t-test being suited best for comparing unimodal distributions. A slight departure from that unimodality, as alluded to earlier, is visible for fake article coherence score. This is most expressed for the ESA method, with a the leftmost red pillar being quite tall in Figures~\ref{fig:esahistisot} and~\ref{fig:esahisthwb}; it may be noted that the leftmost pillars count all documents below that score and thus, the small peak that exists in the lower end of the fake news scores is overemphasized in the graphs due to the nature of the plots. 

Thirdly, the red-green reversal trend as observed earlier, may be interpreted as largely being an artifact of the relative positioning of the centres of the unimodal score distribution across fake and legitimate news articles. The red peak appears to the left (i.e., at a lower coherence score) of the green peak; it is easy to observe this trend if one looks at the red and green distributions separately on the various charts. For example, the red peak for Figure~\ref{fig:wordembhistisot} is at 0.55, whereas the green peak is at 0.60. Similarly, the red peak in Figure~\ref{fig:entityhistisot} is at 0.28, whereas the green peak is at 0.30. Similar trends are easier to observe for the HWB results.

\subsubsection{Summary}

Main observations from our computational social science study are:

\begin{itemize}
\item {\it Fake news articles less coherent:} The trends across all the coherence scoring mechanisms over the two widely different datasets (different domains, different sizes) indicate that fake news articles are less coherent than legitimate news ones. The trends are statistically significant in all but one case. 
\item {\it Word embeddings most suited:} From across our results, we observe that word embedding based mechanism is most suited, among our methods, to help discern the difference in coherence between fake and legitimate news articles. 
\item {\it Unimodal distributions with different peaks:} The high-level trend points to a unimodal distribution of coherence scores (slight departure observed for fake articles), with the score distribution for fake news peaking at a lower score.
\end{itemize}



\section{Discussion}

The relevance or importance of this computational social science study lies in what the observations and the insights from them point us to. We discuss some such aspects in this section. 


\subsection{Towards more generic fake news detection}
\label{sec:generic}

Fake news detection has been traditionally viewed as a classification problem by the data science community, with the early approaches relying on the availability of labelled training data. The text features employed were standard ones used within the text mining/NLP community, such as word-level ones and coarser lexical features. Standard text classification scenarios such as identifying the label to be associated with a scholarly article or determining disease severity from medical reports involve settings where the human actors involved in generating the text are largely passive to the classification task. On the other hand, data science based fake news identification methods stand between the fake news author and accomplishment of his/her objectives (which may be one of myriad possibilities such as influencing the reader's political or medical choices). Social networks regularly use fake news detection algorithms to aid prioritization of stories. Making use of low-level features such as words and lexical patterns in fake news detection makes it easier for the fake news author to circumvent the fake news filter and reach a broad audience. As an example, a fake news filter that is trained on US based political corpora using word-level features could be easily surpassed by using words and phrases that are not commonly used within US politics reporting (e.g., replacing {\it President} with {\it head of state}). On the other hand, moving from low-level lexical features (e.g., words) to higher level ones such as topical novelty (as investigated in a Science journal article~\cite{vosoughi2018spread}), emotions~\cite{emotioncog} and rhetorical structures~\cite{rubin2015towards} would yield more {\it `generic'} fake news identification methods that are more robust to being easily tricked by fake news peddlers. We observe that moving to higher-level features for fake news identification has not yet been a widespread trend within the data science community; this is likely to be due to the fact that investigating specific trends do not necessarily individually improve the state-of-the-art for fake news detection using conventional metrics such as empirical accuracy that are employed within data science. Nonetheless, such efforts yield insights which hold much promise in collectively leading to more generic fake news detection for the future. Techniques that use high-level features may also be better transferable across domains and geographies. We believe that our work investigates an important high-level feature, that of lexical coherence, and provides insights highly supported by datasets across widely varying domains, and would be valuable in contributing to a general fake news detection method less reliant on abundant training datasets. 

\subsection{Lexical Coherence and Other Disciplines}
\label{sec:discussiondisc}

We now consider how our observations could lead to interesting research questions in other disciplinary domains. 


\subsubsection{News Media}

The observation that fake news articles exhibit lower textual coherence could be translated into various questions when it comes to news media. It may be expected that fake news articles appear in media sources that are not as established, big or reputed as the ones that report only accurate news; these media houses may also have less experienced staff. If such an assumption is true, one could possibly consider various media-specific reasons for the relate coherence trends between fake and legitimate news:

\begin{itemize}[leftmargin=*]
\item Is the lower coherence of fake news articles a reflection of less mature media quality control employed within smaller and less established media houses?
\item Is the relative lack of coherence of fake news articles more reflective of amateurish authors with limited journalistic experience?
\end{itemize}

\subsubsection{Linguistics}

Yet another perspective to analyze our results is that of linguistics, or more specifically, cognitive linguistics\footnote{https://en.wikipedia.org/wiki/Cognitive\_linguistics}. The intent of fake news authors are likely more towards swaying the reader towards a particular perspective/stance/position; this is markedly different from just conveying accurate and informative news, as may be expected of more reputed media. Under the lens of cognitive linguistics, it would be interesting to analyze the following questions:

\begin{itemize}[leftmargin=*]
\item Is the reduced coherence due to fake news generally delving into multiple topics within the same article?
\item Could the reduced coherence be correlated with mixing of emotional and factual narratives?
\end{itemize}

\noindent Besides the above, our results may spawn other questions based on perspectives that we are unable to view it from, given the limited breadth in our expertise.

\section{Conclusions and Future Work}

We studied the variation between fake and legitimate news articles in terms of the coherence of their textual content. Within the computational social science framework, we used state-of-the-art data science methodologies to formulate three different computational notions of textual coherence, in order to address our research question. The methods widely vary in character and use word embeddings, explicit semantic analysis and entity linking respectively. We empirically analyzed two datasets, one public dataset from politics, and another one comprising health and well-being articles, in terms of their textual coherence, and analyzed differences across their fake and legitimate news subsets. All the results across all six combinations (3 scoring methods, 2 datasets) unequivocally indicated that fake news article subset exhibits lower textual coherence as compared to the legitimate news subset. These results are despite the fact that fake news articles were found to be shorter, short articles intuitively having a higher propensity to be more coherent. In summary, our results suggest that fake news articles are less coherent than legitimate news articles in systematic and discernible ways when analyzed using simple textual coherence scoring methods such as the one we have devised. 



\subsubsection*{Acknowledgements:} Deepak P was supported by MHRD SPARC (P620). 

%
%
%
\bibliographystyle{splncs04}
\footnotesize
\bibliography{research}

\end{document}